\documentclass[twocolumn,letterpaper]{IEEEAerospaceCLS} 

\usepackage{cite} 
\usepackage{amsmath,amssymb,amsfonts}
\usepackage{algorithmic}
\usepackage{graphicx}
\usepackage{textcomp}
\usepackage{xcolor}
\def\BibTeX{{\rm B\kern-.05em{\sc i\kern-.025em b}\kern-.08em
    T\kern-.1667em\lower.7ex\hbox{E}\kern-.125emX}}

\begin{document}

\title{Space Trusted Autonomy Readiness Levels\thanks{Approved for public release; distribution is unlimited. Public Affairs approval numbers: AFRL-2022-4079, PAIRS CASE 2022-0638, NRO 2022-04386, MITRE PRS Approval Case \# 22-3106, Aerospace OTR202201088.}}

\author{%
Kerianne L. Hobbs\\ 
Air Force Research Laboratory\\
2241 Avionics Circle\\
Wright-Patterson AFB, OH, 45433\\
\and 
Joseph B. Lyons \\
711 Human Performance Wing \\
2210 8th St \\
Wright-Patterson AFB OH 45433 \\
\and 
Martin S. Feather\\
Jet Propulsion Laboratory, California Institute of Technology\\
4800 Oak Grove Dr. \\
Pasadena, CA 91109\\
\and 
Benjamen P Bycroft\\
The Aerospace Corporation\\
2310 E. El Segundo Blvd \\
El Segundo, CA 90245\\
\and 
Sean Phillips\\
Air Force Research Laboratory\\
3550 Aberdeen Ave SE \\
Kirtland AFB, NM, 87117
\and 
Michelle Simon\\
Air Force Research Laboratory\\
3550 Aberdeen Ave SE \\
Kirtland AFB, NM, 87117\\
\and 
Mark Harter\\
The MITRE Corporation \\
1155 Academy Park Loop \\
Colorado Springs CO 80910\\
\and 
Kenneth Costello\\
NASA\\
100 University Dr\\
Fairmont, WV  26554 \\
\and 
Yuri Gawdiak\\
NASA Headquarters \\
Washington DC, 20546 \\\
\and
Stephen Paine\\
National Reconnaissance Office\\
14675 Lee Rd\\
Chantilly, VA 20151\\
\thanks{{U.S. Government work not protected by U.S. copyright; C2022: The MITRE Corporation. All Rights Reserved.}}         
}

\maketitle

\thispagestyle{plain}
\pagestyle{plain}

\begin{abstract}

Technology Readiness Levels are a mainstay for organizations that fund, develop, test, acquire, or use technologies. Technology Readiness Levels provide a standardized assessment of a technology’s maturity and enable consistent comparison among technologies. They inform decisions throughout a technology’s development life cycle, from concept, through development, to use. A variety of alternative Readiness Levels have been developed, including Algorithm Readiness Levels, Manufacturing Readiness Levels, Human Readiness Levels, Commercialization Readiness Levels, Machine Learning Readiness Levels, and Technology Commitment Levels. However, while Technology Readiness Levels have been increasingly applied to emerging disciplines, there are unique challenges to assessing the rapidly developing capabilities of autonomy. This paper adopts the moniker of Space Trusted Autonomy Readiness Levels to identify a two-dimensional scale of readiness and trust appropriate for the special challenges of assessing autonomy technologies that seek space use. It draws inspiration from other readiness levels’ definitions, and from the rich field of trust and trustworthiness. The Space Trusted Autonomy Readiness Levels were developed by a collaborative Space Trusted Autonomy subgroup, which was created from The Space Science and Technology Partnership Forum between the United States Space Force, the National Aeronautics and Space Administration, and the National Reconnaissance Office.

\end{abstract}


\tableofcontents

\section{Introduction}

Technology Readiness Levels (TRLs) were initially developed by NASA in the 1970s to assess the maturity of technologies with application to space \cite{sadin1989nasa}. The final 1-9 scale, with 1 being the least mature and 9 being the most mature, was formalized by NASA in the 1990s \cite{mankins1995technology}. The role of TRLs in Aeronautics was discussed in \cite{dunbar_2015} and a NASA Best Practices Guide provided in 2020 \cite{beauchamp2020technology}. TRLs have spread to become a mainstay in United States Air Force Acquisition processes\cite{AcqNotes2021TRL} and the European Space Agency \cite{heder2017nasa} as well as across several other agencies. Although the TRL scale has increasingly been applied across disciplines well outside its initial space origin, researchers have argued that its meaning diminishes and causes confusion \cite{heder2017nasa}. Perhaps for this reason, a variety of alternative readiness levels (RLs) have been developed, including Algorithm RLs (ARLs) \cite{Bateman2016}, Manufacturing RLs (MRLs) \cite{morgan2015}, Human RLs (HRLs) \cite{ANSI_HFES400_2021}, Commercialization RLs (CRLs)\cite{Maybury2014}, Data RLs (DRLs)\cite{lawrence2017data}, Machine Learning RLs (MLRLs)  \cite{lavin2021technology}, Transition Commitment Levels (TCLs) \cite{TCLs} and others.

Space trusted autonomy lies at the intersection of traditional aerospace engineering, computer and data science, human-autonomy interaction, and multi-system interaction, which makes the application of traditional TRLs problematic for the following reasons.  
First, assessment of autonomy readiness hinges on measuring intangible capabilities. For non-autonomous technologies destined for space applications, tangible factors such as Size, Weight and Power (SWaP) are more readily tracked through maturity levels and performance metrics. Autonomy technologies have informational needs – dependencies on data, signals, commands, and the frequency and quality of such information. Their capabilities are primarily decision and outcome based, rather than measured in terms of physical parameters. Assessment of autonomy is made more challenging by the requirement to operate appropriately in a wide and often poorly circumscribed range of circumstances. 
Second, autonomy represents a transfer of delegated and bounded authority from humans to the autonomous system, and thus necessitates including trust as an entire additional dimension of readiness assessment. Previous TRLs assume human controllers, who have values, ethics, and context beyond engineered controllers. There is an implicit trustworthiness in human operators that doesn't transfer to autonomy directly. Trust is a fundamental social process wherein a trustor evaluates the trustworthiness of a referent and makes a decision (or series of decisions) related to willingness to be vulnerable to that referent. For space use, there are multiple trustor communities – the proponents of the technology, the engineers and testers instrumental to its development, the operators of the space mission who will decide when and how much to deploy the technology and interact with it, and the consumers who seek the results achieved through its use. Third, assessing space autonomy readiness is challenging because it's impossible to completely replicate space conditions terrestrially (e.g., radiation, gravity, dust, etc.) and testing autonomy on orbit comes at high expense and high consequences of mistakes (e.g., a collision that cascades into more collisions, creating significant amount of space debris), which discourages deploying a prototype in space purely for testing purposes.

The Space Science and Technology Partnership Forum between the United States Space Force (USSF), the National Aeronautics and Space
Administration (NASA), and the National Reconnaissance Office (NRO) created the Space Trusted Autonomy subgroup in 2019 to identify synergistic opportunities across the U.S. Government. The partnership defined \cite{jones2021recommendations} \textit{space} as ``focused on space systems," \textit{trusted} as behavior in which human operators and stakeholders have confidence, and \textit{autonomy} as ``some level of decision-making authority that resides within the system" (for comparison, Merriam-Webster defines autonomy as ``having the right or power of self-government" or ``undertaken or carried on without outside control" \cite{MWAutonomous}). Trust is not the same thing as trustworthiness, which is defined as ``the real competency of a system to perform functions given the extent of the authority it has been granted and the consequences of its potential actions" \cite{lacher2017framework}.
This manuscript reports the partnership’s efforts to develop ``Space Trusted Autonomy Readiness" (STAR) Levels to serve as readiness levels in this domain.


\section{Autonomy}
Merriam-Webster defines autonomy as ``the quality or state of being self-governing" and autonomous as ``a response undertaken or carried on without outside control; responding, reacting, or developing independently of the whole"  \cite{MWAutonomous}. 
%
Autonomy has been essential to space exploration, and is becoming increasingly vital to all aspects of modern and future space operations. Opportunities for space autonomy include critical maneuvers such as Entry, Descent and Landing (EDL), targeting of science instruments during short-duration fly-bys, and reacting to internal spacecraft faults to preserve the fundamental health of the spacecraft while it communicates to Earth for guidance on how to recover and proceed. Autonomy is necessary in missions where prohibitive communication delays due to extreme distances and occultation require on-board capabilities to perceive, decide upon the course of action, and execute. Autonomy has also been used to enhance the accomplishment of science. For example, NASA’s rovers on Mars use autonomy to drive long distances by themselves, to recognize the occurrence of transient phenomena so as to gather and return data on those, and to select science targets for investigation. As another example, the USSF/SPOC is interested in autonomy capabilities to enable the rapid proliferation of LEO constellations (P-LEO), such as autonomy in Space Surveillance Network (SSN) sensor tasking, tracking satellites, Space Domain Awareness (SDA), and Space Traffic Management (STM) for mission assurance, safety of flight, and protection of public space systems.  
In the closer regions of cislunar space, the exponential increase in the number of space objects is making it infeasible to recruit and train human operators at the traditional ratio of a control room of operators per spacecraft. Continued proliferation of space systems rapidly outpaces the existence of ground-station coverage and availability. This too is driving the development and deployment of autonomy, to aid in the management, supervisory control, and operations of spacecraft and thus allow a small number of operators to manage a large number of spacecraft using finite ground-station resources.

Autonomy in these settings comprises the independent decision-making made by space assets given bounded and delegated authority from human operators. A space asset could be an individual subsystem, an entire spacecraft, a planetary lander or rover, etc., or a system comprising multiples of these. Autonomous decisions will influence the asset’s actions, e.g., directing maneuvering, powering systems on and off, guiding what observations to perform, or selecting what data to send to Earth.
%

\subsection{Autonomy Levels}
Conversations about the technical maturity of autonomous space systems inevitably lead to a discussion about levels of autonomy themselves. A number of autonomy levels have been defined by the research community, and these differ from readiness levels. Autonomy levels usually define specific roles for humans versus automation/autonomy, while readiness levels define the technical maturity of those systems. Additionally, there are several definitions of automation and autonomy, such as that by NASA where automation ``is the automatically controlled operation of an apparatus, process, or system using a pre-planned set of instructions (e.g., a command sequence)," and autonomy is defined as ``is the capacity of a system to achieve goals while operating independently from external control" \cite{miller2015nasa}. A brief summary of autonomy levels follows so that the reader will appreciate how they are relevant to understanding the scope of a system’s autonomy, but do not themselves address the focus of this paper, namely issues of trust and readiness.


The Society of Automation Engineers (SAE) defined a 0 to 5 scale for levels of driving automation based on the roles of the human versus the automation. Levels 0-2 feature the human in primary control using driver support features like warnings (automatic emergency braking, blind spot, lane departure), brake/acceleration support, and features like lane centering and/or adaptive cruise control. In Levels 3-4 the automation is in primary control in limited operational conditions. At level 3, the automation may still request that the human take over, but by level 4 the car may no longer have a steering wheel or pedals. By level 5, the car can safely drive under all conditions.

Aviation Autonomy Levels \cite{anderson2018levels} have been defined on a similar 0 to 5 scale, as follows: 0 (no automation), 1 (assistance), 2 (partial automation), 3 (highly automated) 4 (fully automated), and 5 (autonomous).  These levels were further defined relative to specific pilot tasks: aviate (maintain control and avoid collisions), navigate (identify current location and flight plan), communicate (coordinate with other systems and traffic controllers), manage systems (configure for different flight phases as well as detect, identify, and mitigate faults), and make command-level (tactical and strategic operational) decisions.


\subsection{Qualification, Certification, and Flight Proven}
The concept of TRLs is also closely related to that of evidence-supported qualification, certification, and flight proven. \textit{Evidence} is data created through the development process that supports the use of the system, may include a formal certification, and is also integral to engendering trust in the system.
The following definitions of Qualification and Certification are from NASA’s Systems Engineering Handbook \cite{hirshorn2017nasa}. \textit{Qualification} activities are performed to ensure that the flight unit design will meet functional and performance requirements in anticipated environmental conditions. A subset of the verification program is performed at the extremes of the environmental envelope and will ensure the design will operate properly with the expected margins. Qualification is performed once regardless of how many flight units may be generated (as long as the design does not change). \textit{Certification} is the audit process by which the body of evidence that results from the verification activities and other activities are provided to the appropriate certifying authority to indicate the design is certified for flight/use. The Certification activity is performed once regardless of how many flight units may be generated. Actual system \textit{flight proven} through successful mission operations (from Table 2.3.2-1 TRL Definition and Decomposition by Factor \cite{beauchamp2020technology}):
\begin{itemize}
    \item Completion criteria: Documented mission operational results verifying requirements.
    \item Performance/Function: Required functionality/ performance demonstrated.
    \item Fidelity of Build: Final product: Flight unit.
    \item Environment Verification: Operated in actual operational environment.
\end{itemize}

\section{Trust}

A seminal review on trust in automation published by See and Lee in 2004 \cite{lee2004trust}, and later updated in a survey paper by Hoff and Bashir in 2015 \cite{hoff2015trust}, define trust as ``the attitude that an agent will help achieve an individual’s goals in a situation characterized by uncertainty and vulnerability" \cite{lee2004trust}. Trust relies on a gain for individual's goals or mission that justify taking a risk. Trust is a fundamental social process wherein a trustor evaluates the trustworthiness of a referent and makes a decision (or series of decisions) related to how willing she or he is to be vulnerable to that referent \cite{mayer1995integrative}. Trust referents can be other humans (or organizations of humans \cite{mayer1995integrative}) or technological systems. Understanding human trust in autonomy is critical because trust influences reliance behavior \cite{lee2004trust}, yet there are many challenges associated with the trust process. One of the biggest challenges is ensuring that operator trust of space autonomy is appropriately calibrated. The risk of overtrust is automation misuse when “people inadvertently violate critical assumptions and rely on automation” in situations where it is ill suited. This overtrust has been shown to be associated with severe performance degradation when the human is using a faulty automation aid \cite{onnasch2014human}. Since all automation should be considered imperfect, just like humans, the issue of establishing and maintaining effective trust calibration is paramount to the success of automated systems. The risk of undertrust is automation disuse which occurs “when people reject the capabilities of automation.” Not using an otherwise reliable system can result in inefficiencies and higher costs, thus neither overtrust nor undertrust are desired for the fielding of novel automation. These challenges are amplified in the space domain and the next section highlights key insights from the trust in automation research community applicable to Space Trusted Autonomy RLs.

\subsection{Trust Attributes and Progression}

Trust is dynamic \cite{audrey2018conceptualising,falcone2004trust} rather than static and influenced by many different factors. Interpersonal trust between two or more humans involves an evaluation of the referents' trustworthiness, operationalized by Mayer and colleagues as ability, benevolence, and integrity. \textit{Ability} corresponds to one's task competence. \textit{Benevolence} refers to the belief that the referent will act in support of the trustor's goals. \textit{Integrity} represents the belief that the referent possesses a stable set of values that are acceptable to the trustor. Meta-analytical research has found that ability, benevolence, and integrity each account for unique variance in predicting trust \cite{de2016trust}. When the trust referent is a machine (such as automation) versus a human, there are both similarities and differences in terms of what factors shape trust.   

In the human-automation trust research, humans tend to start by assuming automation is perfect, resulting in overtrust that erodes quickly with any error. This is referred to as a perfect automation schema. When considering automation, humans view the automation as more invariant and performance-oriented. They are more sensitive to errors compared to a human referent \cite{madhavan2007similarities}. Thus, performance will certainly be a major factor in shaping trust readiness levels. Performance of automation (also termed reliability) has consistently been shown to relate to trust of the automation \cite{hoff2015trust}. These effects have been found in laboratory contexts as well as studies of automation in the field. For instance, in studies of pilot trust in automation, nuisance avoidance (a performance element) was a prominent factor for developing trust \cite{ho2017trust}. One pilot in the study indicated that one mistake by the automation would likely cause pilots to turn the system off and lose the advantage it provided, which tracks with other similar research that found frequent false-alarm rates lead pilots to deactivate critical alarm systems \cite{sorkin1988people}. Yet, despite the importance of performance, one's trust of automation is driven by a number of factors. In fact, contemporary thoughts on trust in automation have begun to emphasize relational attributes, transparency of the automation, and responsiveness/resilience of automation across contexts \cite{chiou2021trusting, calhoun2019linking,chen2020guest}.

Equally important to trust is the operator’s understanding of the automated system’s methods (process-based trust), and the design intent of the autonomous system (purpose-based trust). The operator needs a mental model of the automated system and the automated system may need a model of the operator. For example, in aircraft automation cases, it has been found that simplifying algorithms so that they are understandable to the end user is an important factor in calibrating user trust \cite{lee2004trust}, and that familiarity with the automated behavior and its consistency with the operator’s training and manual approaches resulted in strong positive perceptions of the system \cite{lyons2016trust}. A key element of this literature is the construct of transparency. Transparency methods could highlight the decision rationale used by an automated tool \cite{lyons2016trust}; facilitate perception, understanding, and projection of an agent's behavior; or highlight the intent-/team-based knowledge of an interaction \cite{chen2020guest,mercado2016intelligent}. Thus, understanding, like performance, will be a key feature of the trust readiness level.

\subsection{Human-Automation Trust Variability}

Human trust in automation is not universal and variability has been categorized by an individual’s propensity to trust automation (dispositional trust), the specific context of the automation interaction (situational trust) and a human’s past experiences with the specific automated system (learned trust) \cite{marsh2003role}. In terms of dispositional trust, very limited research indicates that \textit{culture} (e.g., Mexicans trust automated decision aids more than Americans \cite{huerta2012framing}), \textit{age} (e.g., older adults are more likely to trust \cite{ho2005age}, and have better calibrated trust to decision aids\cite{sanchez2004reliability}, and were less likely to be swayed by the picture of a human expert in the interface than younger adults\cite{pak2012decision}), \textit{gender} (e.g., women respond more positively to flattery by automation than men \cite{lee2008flattery}, and respond differently to automated system communication styles and appearances \cite{nomura2008prediction,tung2011influence}), and \textit{personality} (e.g., neurotic people are less likely to trust automation \cite{szalma2011individual}, extroverts are more likely to trust automation \cite{merritt2008not}, intuitive personalities are more trusting than sensing personalities \cite{mcbride2012impact} (not to suggest that intuitive personalities are more gullible, but rather how they ingest and process information about the automated system makes them trust it more), and some people just trust automation more than others \cite{colquitt2007trust}). However, much more research is needed to identify consistent patterns.
\textit{Situational} trust depends on context, environment, and the operator's mental state \cite{marsh2003role}. For example, when an operator has a very high workload, they are likely to rely on the automated assistant even if they don't trust it \cite{biros2004influence}.
\textit{Learned} trust is earned through operator's experience with the automated system over time \cite{marsh2003role}. While users may have an initial level of trust in an automated system, the trust level changes dynamically throughout the user's experience \cite{hoff2015trust}. Users may also initially trust a system more if the automated system has a good reputation, but may trust it less if the user is a subject matter expert in the process that has been automated \cite{hoff2015trust}.

\subsection{Different Levels of Risk Tolerance}
\begin{figure}
    \centering
    \includegraphics[width = .49\textwidth]{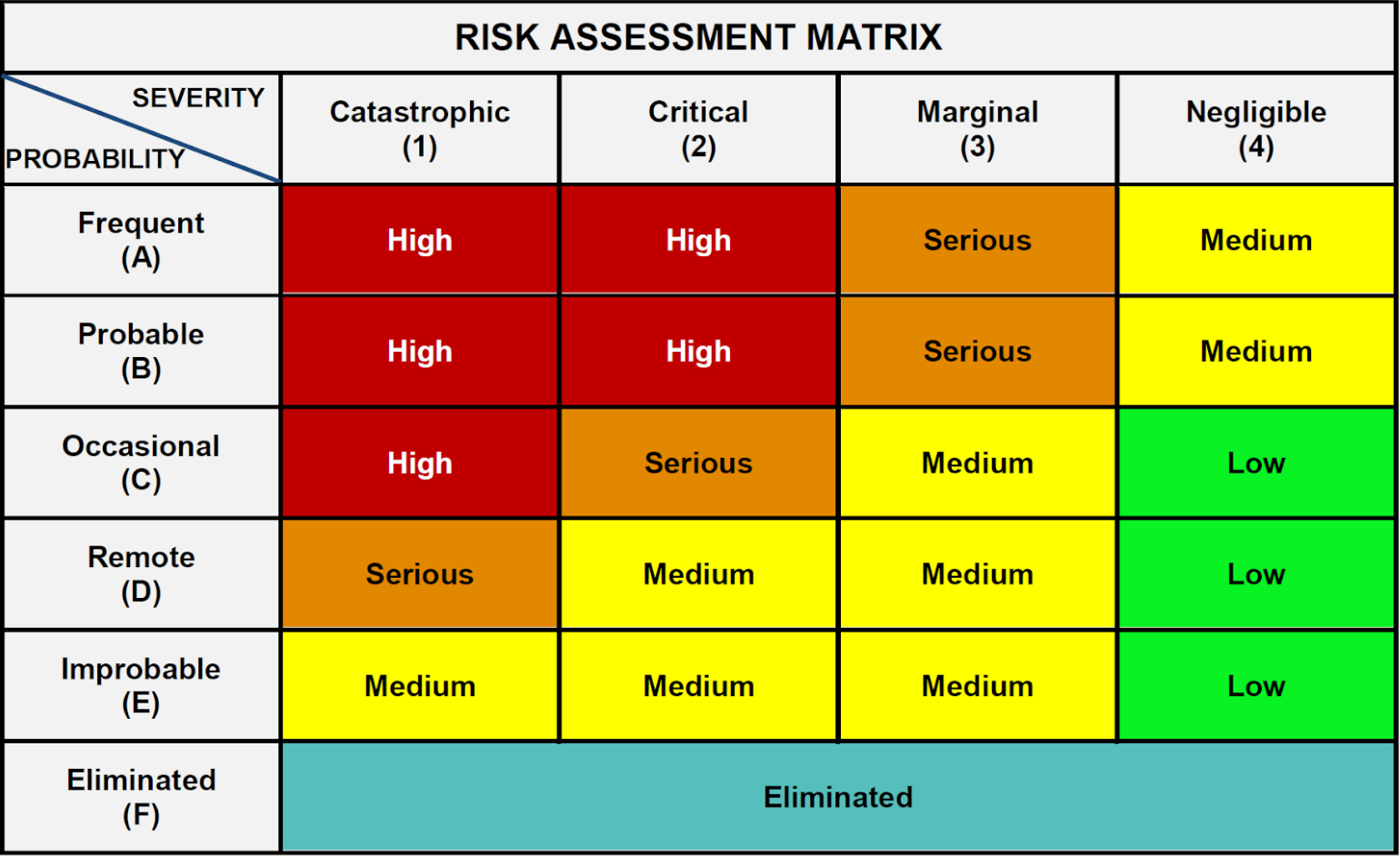}
    \caption{Military Risk Assessment Matrix \cite{MILSTD882E}}
    \label{fig:MIL-STD-882E}
\end{figure}

\begin{figure}
    \centering
    \includegraphics[width = .4\textwidth]{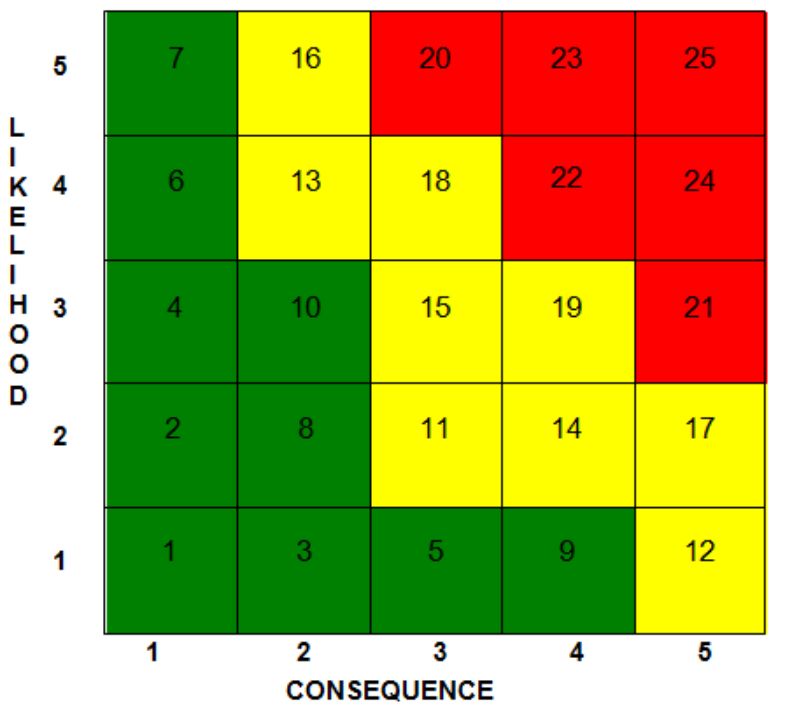}
    \caption{NASA Risk Assessment Matrix \cite{S3001}}
    \label{fig:S3001Risk Matrix}
\end{figure}

Trust, when defined as a willingness to be vulnerable, is analogous to a willingness to accept risk. Risk is defined by the military in terms of probability (or likelihood) and severity (or consequence). Military and NASA Definitions of probability are presented in Table \ref{tab:probability}. On the NASA side, consequence is specified in 5 levels for each of performance, human safety, asset, schedule, and cost consequences \cite{S3001}. Military severity categories are defined in terms of human safety and cost, and are summarized in Table \ref{tab:severity} \cite{MILSTD882E}. The combination of probability and severity is then organized in a matrix and assigned a risk level. The military risk matrix is shown in Figure \ref{fig:MIL-STD-882E} and the NASA risk matrix is shown in Figure \ref{fig:S3001Risk Matrix}. It is worth noting that the Military standard also defines software safety criticality based on severity and level of software control (including levels of no safety impact, influential, redundant fault tolerant, semi-autonomous, and autonomous control).
\begin{table*}[]
    \centering
    \caption{Probability Categories from MIL-STD-882E \cite{MILSTD882E} and NASA S3001 \cite{S3001}}
    \begin{tabular}{|p{1cm}|p{8cm}|p{1cm}|p{2.7cm}||p{2.5cm}|}\hline
        882E Level  & 882E Criteria & S3001 Level & S3001 Criteria & S3001 Probability \\\hline
         A & Likely to occur often in the life of an item. &5 & Near certainty &$p>80$\%\\\hline
         B & Will occur several times in the life of an item.&4 & Highly Likely &80\%$>p>60$\%\\\hline
         C & Likely to occur sometime in the life of an item. &3& Likely  &60\%$>p>40$\% \\\hline
         D & Unlikely, but possible to occur in the life of an item. & 2 & Low likelihood &40\%$>p>20$\%\\\hline
         E & So unlikely, it can be assumed occurrence may not be experienced in the life of an item. &1 &Not likely &20\%$>p>0$\%\\\hline
         F & Incapable of occurrence. This level is used when potential hazards are identified and later eliminated.& & & \\\hline
    \end{tabular}
    \label{tab:probability}
\end{table*}
\begin{table}[]
    \centering
    \caption{Severity categories used to assign risk under MIL-STD-882E \cite{MILSTD882E}}
    \begin{tabular}{|p{1.7cm}|p{1cm}|p{4.7cm}|}\hline
         Description & Severity & Mishap Result Criteria \\\hline
         Catastrophic & 1 & Death, permanent total disability, irreversible significant environmental impact, or monetary loss $\geq$ \$10M. \\\hline
         Critical & 2 & Permanent partial disability, injuries or occupational illness that may result in hospitalization of at least three personnel, reversible significant environmental impact, or monetary $\geq$\$1M and $<$ \$10M. \\\hline
         Marginal & 3&  Injury or occupational illness resulting in one or more lost work day(s), reversible moderate environmental impact, or monetary loss $\geq$ \$100K and $<$ \$1M. \\\hline
         Negligible & 4 & Injury or occupational illness not resulting in a lost work day, minimal environmental impact, or monetary $<$ \$100K.\\\hline
    \end{tabular}
    \label{tab:severity}
\end{table}
One challenge in development of operational automation and autonomy is that different factors such as risk tolerance and resource margins (e.g., of propellant) may differ between missions or operators. For example, two different space missions may have different risk tolerances for mission failure. One way to deal with this is to offer multiple modes that the operator can switch between that offer varying safety buffers \cite{swihart2011automatic, swihart2011design}. Another option is to apply more rigorous verification and validation to a system for an application with less tolerance for risk.

\subsection{Accountability, Vulnerability, and Expectations}
Studies in the design of aerospace automation have found that mutual understanding and respect between operators, engineers, and program managers that feature different accountability, vulnerability, and expectations is important to success \cite{ho2017trust}. Operators, as the end users accountable for successful missions, are vulnerable to failure, and expect the automation to perform acceptably. Engineers, accountable for designing the system to perform well, are vulnerable to backlash from managers and operators if the system is poorly designed, and expect the system to be used as the designer intended. Managers, accountable for accomplishing larger mission goals for the system, are vulnerable to loss of resources to complete system design and evaluation, and expect the system to meet mission needs.  Mutual understanding of these different accountabilities, vulnerabilities, and expectations resulted in human-to-human trust that led to greater trust in an automated system \cite{ho2017trust}. 

A MITRE study extended this list of stakeholders to a researcher (reputation at risk), a regulator (reputation, job security and public trust at risk), a creator implementing the design (reputation, job security, investor's finances at risk), insurers (job security, company finances at risk), the acquirer (mission effectiveness, organization finances, and job security at risk), the commander or supervisor (job security, mission effectiveness, personal safety, personal finances at risk), the operator (job security, mission effectiveness, personal safety and finances at risk), the patron (personal finances and property at risk), and the community (personal safety, property and finances at risk) \cite{lacher2014autonomy}. 

In the arena of space missions whose purpose is science-driven, the Framework for Trusted AI developed by The Aerospace Corporation, when applied to an example of AI-based autonomy, identified two additional stakeholder communities \cite{mandrake2022space}. These were the Science Team – those seeking to address the mission’s science goals and objectives, and the broader Science Community – the ultimate end-users of the mission’s data. Regardless of the trust built up in the mission team, any scientific conclusions drawn from the returned data must withstand skeptical scrutiny. If autonomy is present in the data flow itself, it is crucial that it be trusted to not produce erroneous results or induce biases that could call into question analysis results.



\subsection{Ethics and Trust}
Programming and implementing automation, autonomy, and AI to align with ethical principles is an important facet of trust. Ethics may be defined as ``a set of reflected norms, rules, precepts, and principles that govern and guide the behavior of individuals or groups" \cite{Pflanzer2022Ethics}. As an example of how ethics might be applied to automation, autonomy, and AI, the U.S. Department of Defense has adopted ethical principles that AI be responsible, equitable, traceable, reliable, and governable \cite{DOD2020EthicalAI}. The Intelligence Community also published principles of artificial intelligence ethics which include: respecting the law and acting with integrity, providing transparency and being accountable for AI use and its outcomes, using objective and equitable AI that mitigates bias, developing and using AI centered around humans, ensuring secure and resilient AI lifecycle, and applying rigorous science and technology practices in the development and use of AI \cite{ODNI2020EthicalAI}.


\subsection{Incremental Trust Gains During Operations}

Some deployments of autonomy offer opportunities to incrementally gain trust during operations through experiential data/information. 
One approach is to let the autonomy decide what it would do, but have it stop short of actually carrying out its action. This allows operators to confirm the correctness of its operation while avoiding the risk of the autonomy incorrectly performing a dangerous action. Autonomy in this sense can support the human in information acquisition and analysis while leaving the action implementation and decision making up to the human. When the task temporal demands allow for this flexibility it can be a robust way to incorporate autonomy into a task context. This does leave the risk of the autonomy software misbehaving in some way, e.g., interfering with other critical processes or corrupting critical data, however such risks can generally be assuaged by good software engineering practices. 
This incremental approach was followed by activation of the AEGIS system discussed earlier \cite{estlin2012aegis}, in which it was allowed to exercise an increasing number of steps to plan a science gathering activity, and develop confidence before finally allowing AEGIS to actually perform that activity.

If the space system can recognize when an activity is heading toward danger in time to safely terminate that action, then operators will have confidence to let the autonomy operate. This is the approach taken by NASA’s rovers on Mars, where on-board software safeguards the rover during all activities. For the Curiosity rover, \cite{rankin2020driving} describes how during driving, “mobility fault protection” evaluates the rover state at 8Hz, stopping the drive if any state relevant to driving is outside of set limits (e.g., rover tilt limit exceeded).

When it is safe to do so, operators can gain confidence in autonomy by engaging it in an increasingly ambitious manner. This was done by ESA (European Space Agency) in field testing an autonomous rover in the Atacama Desert in Chile in preparation for a robotic mission to Mars. \cite{woods2014demonstrating} describes how, over several days of trials during which the autonomy performed as expected, the operations team learned to trust the autonomy components and switched to using high-level autonomy tasks to perform increasingly complex operations.

\subsubsection{Incremental Trust Gains from Live-Virtual-Constructive Simulation and Test}
A relatively new concept in the space community to build incremental trust in cooperative spacecraft autonomy is borrowed from the Unmanned Aerial Systems (UAS) Community: Live-Virtual-Constructive (LVC) testing. In this testing approach a \textit{constructive} spacecraft is one that exists only in a computer simulation, a \textit{virtual} spacecraft is one that might be implemented in physical simulator on the ground (e.g., an air-bearing attitude determination and control test bed, a surrogate test bed like an omnicopter), and a \textit{live} simulator is an actual spacecraft operating in space that is performing autonomous maneuvering. In this build up approach, two cooperative satellites might first be entirely simulated to validate the initial concept. Then, a virtual test bed might be used to demonstrate interactions. Then combinations of a live spacecraft and constructive or virtual can be used to validate how a real spacecraft in a space environment would behave relative to a constructive or virtual asset simulated on the ground (but the live agent believes also exists in orbit). Next, a live with live-offset can be used where two spacecraft are tested with a safe separation, but believing they are closer together (e.g. simulating proximity operators at 1 km apart when the spacecraft are actually 50 km apart). Then finally, when enough confidence is gained, the spacecraft can operate in close proximity to one another.
\section{Space Trusted Autonomy Readiness Levels}
\begin{figure}
    \centering
    \includegraphics[width = .5\textwidth]{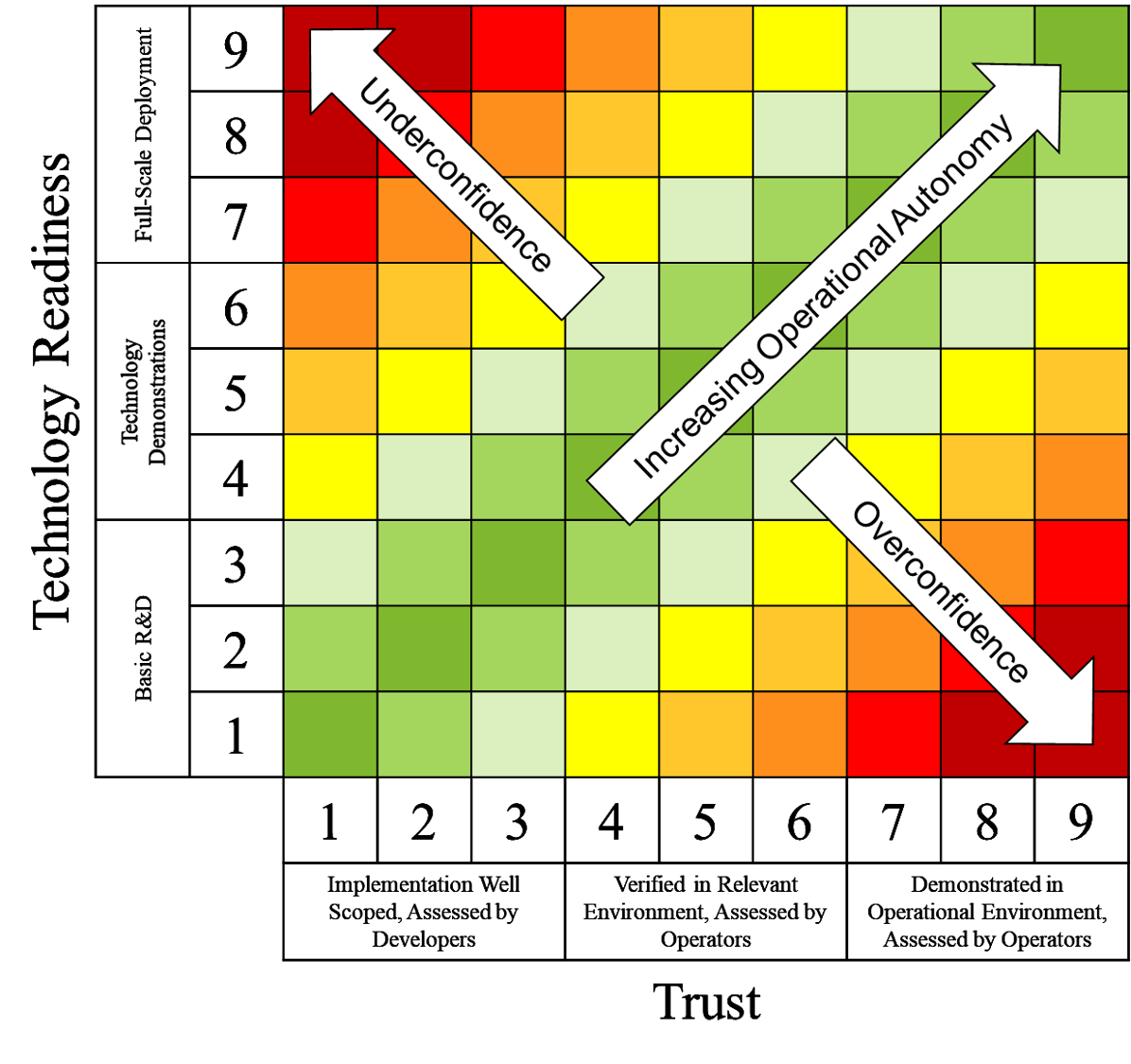}
    \caption{Trust vs Readiness, Annotated}
    \label{fig:readinessmatrix}
\end{figure}

In this section, STAR Levels are presented as orthogonal axes of readiness: autonomous capability and trust. The need to balance readiness levels, while important to any program \cite{vik2021balanced}, becomes acute in this two-dimensional representation of readiness.  As depicted in Figure \ref{fig:readinessmatrix}, high trust and low readiness is associated with over-confidence of the autonomous operations, while low trust and high readiness is associated with under-confidence. These STAR Levels are not intended to replace TRLs, but rather to highlight specific definitions for autonomy which differ from other systems. 

\subsection{Autonomy Readiness Levels}
To define autonomy readiness levels, the team examined several other readiness levels including ARLs \cite{Bateman2016}, MRLs \cite{morgan2015}, HRLs \cite{ANSI_HFES400_2021}, CRLs\cite{Maybury2014}, DRLs\cite{lawrence2017data}, and MLRLs \cite{lavin2021technology}, and TCLs \cite{TCLs}. Then the team started producing their own suggestions, and each was evaluated for commonality. Comparison of these other readiness levels may be found in appendix 2.


The definition of each STAR Level is characterized in terms of the following aspects of maturity:
\begin{enumerate}
\item[{A – Assurance:}] The extent to which the autonomy has been shown to perform correctly.
\item[{C - Context:}] The maturity of the role the autonomy is to play and its embodiment in its intended environmental context.
\item[{I - Implementation:}] The maturity of the autonomy’s implementation (typically in software, but could involve firmware).
\item[{O - Operations:}] The maturity of establishing the interplay between the autonomy and its operators.
\end{enumerate}

STAR Level 1
\begin{enumerate}
\item[{A:}] Some risks or limitations are considered.
\item[{C:}] An opportunity has been identified for autonomy to enhance or enable a space asset’s control or information processing. 
\item[{I:}] A capability in a general or mathematical form is proposed, but no code has been developed.
\item[{O:}] An idea of the potential for an end user to delegate bounded authority has been identified.
\end{enumerate}

STAR Level 2
\begin{enumerate}
\item[{A:}] Performance measures and safety concerns are identified.
\item[{C:}] Initial proof of concept simulations are conducted on simplified models (e.g., point masses, decoupled translational and attitude dynamics) or idealized data (e.g., free of noise or dropouts).
\item[{I:}] Any code or interface is research grade an intended to be used solely by researchers.
\item[{O:}] User concerns are identified.
\end{enumerate}

STAR Level 3
\begin{enumerate}
\item[{A:}] Basic operation of the algorithms demonstrated.
\item[{C:}] Evaluations of core functionality have been performed in low fidelity simulations.
\item[{I:}] Algorithms are maturing and early prototypes of their software implementation have been coded.
\item[{O:}] Human-autonomy and autonomous agent-agent interfaces have been exercised in at least a walk-through manner sufficient to lead to interface requirements and to validate a block-diagram level architecture.
\end{enumerate}

STAR Level 4
\begin{enumerate}
\item[{A:}] The number of scenarios evaluated is relatively small but sufficient to characterize the algorithms’ reliability, safety, and ethical use.
\item[{C:}] Algorithms integrated in a simulation environment for non-real-time test and experimentation.
\item[{I:}] All algorithms are coded in prototype form.
\item[{O:}] Human-autonomy and autonomous agent-agent interfaces are included in prototype
\end{enumerate}

STAR Level 5
\begin{enumerate}
\item[{A:}] The autonomy algorithms are tested to demonstrate proof of concept operations in real time with highly realistic inputs. For example, they could be integrated on spacecraft surrogates, such as spacecraft attitude testbeds, quadcopters, omni-directional drones, air bearing or granite tables, to simulate spacecraft dynamics; alternately, they could be fed playbacks of data recorded on actual missions.
\item[{C:}] Needed capacities are firmly established on computational performance, memory footprint, and input data rates, sufficient for the autonomy to achieve its responsiveness requirements.
\item[{I:}] Prototypes of all algorithms completed.
\item[{O:}] A human-autonomy interface provides operators with insight into the autonomy’s perceptions and decision making processes, based on which they can confirm the autonomy’s behavior, safety and ethics.
\end{enumerate}

STAR Level 6
\begin{enumerate}
\item[{A:}] The autonomy code and interfaces have fully met qualification criteria. Testing has been conducted with both nominal and off-nominal inputs representative of expected conditions of use.  
\item[{C:}] Algorithms and interfaces have been integrated into a relevant computing hardware package, such as a flat sat, that runs within operationally realistic (real-time) constraints.
\item[{I:}] The autonomy code and interfaces have been developed in compliance with all flight-code processes, standards, regulations, and practices.
\item[{O:}] Human-autonomy interface enables operators to understand test results in real time and provide feedback on safety, ethics, and performance.
\end{enumerate}

STAR Level 7
\begin{enumerate}
\item[{A:}] The autonomy is tested for specific use case scenarios and may be wrapped with a run time assurance system that provides safety guarantees and backup control strategies.
\item[{C:}] Autonomy is integrated onto a prototype system and demonstrated in the final use environment, or one that closely mimics relevant environmental characteristics, so that the autonomy is exposed to realistic inputs and called upon to perform in a realistic manner. For example, a live spacecraft on orbit may be flown in parallel to a simulated spacecraft on the ground to evaluate potential scenarios with no risk of collision.
\item[{I:}] Algorithms are maturing and early prototypes of their software implementation have been coded.
\item[{O:}] Autonomy is integrated with a prototype human-autonomy interface, with participation of a multidisciplinary team of aerospace (or similar), quality assurance, safety, and human factors engineers as well as computer scientists. Risks are quantified, and a strategy is defined for how data is obtained, managed, used, secured, and ethically used
\end{enumerate}

STAR Level 8
\begin{enumerate}
\item[{A:}] All necessary formal tests have been successfully passed, sufficient to qualify the autonomy for space use by the identified system in the anticipated operational conditions. Tests included stress testing to cover plausible off-nominal situations.
\item[{C:}] The algorithms have been integrated into the system that will use them.
\item[{I:}] The development of all algorithms and interfaces has been completed, as has their integration into the system that will use them.
\item[{O:}] Tests confirm operations teams’ ability to sufficiently understand the autonomy’s behavior (informed by the feedback it provides) to interact and direct the autonomy as might be needed
\end{enumerate}

STAR Level 9
\begin{enumerate}
\item[{A:}] The autonomous system has been proven through successful mission operations for multiple programs.
\item[{C:}] The autonomy is continuously monitored and evaluated for anomalies in behavior, especially if the autonomous system continues to adapt after deployment.
\item[{I:}] A run time assurance system may be used to monitor specific safety or performance properties and intervene when necessary.
\item[{O:}] Human-autonomy interface meets needs of operators in nominal and off-nominal conditions.
\end{enumerate}

\subsection{Trust Readiness Levels}
While trust can be expressed between two or more autonomous agents, only trust between a human and the autonomous system is considered in these definitions. Several tenets went into these Trust Readiness Levels (TrRLs). Notably, performance and understandability (used as a proxy for the technically-terse term transparency) are key features derived from the literature. An additional three elements were incorporated into the TrRLs. First, the TrRLs should consider perspectives starting with the designer, moving to the tester, and closing with the operator. While operator perspectives should be incorporated from the start of the design and throughout the testing process, direct evaluation with operators is needed to move beyond a TrRL of 5. The second tenet of the TrRLs is the idea that trust should be considered (i.e., shared) across the community of operators. A key aspect of this tenet is that for the highest level of TrRL, the technology must be evaluated across the operational community. Finally, the TrRL must also evaluate the technology across a range of contextual scenarios involving not only nominal situations, but also known contexts wherein errors are possible/probable. Operators must observe the automation in a wide range of contexts in order to appropriately achieve a TrRL of 9. The current TrRL definitions are shown below, and it it is anticipated that additional R\&D is needed to explore and refine the criteria. 

\begin{enumerate}
    \item[{[TrRL1]}] The system's conceptual performance is acceptable to the designer. 
    \item[{[TrRL2]}] The system's task performance is \textit{understandable (traceable and logical)} to the designer.
    \item[{[TrRL3]}] The system's task performance is acceptable and understandable (traceable and logical) to a \textit{tester}.
    \item[{[TrRL4]}] The system's task performance is acceptable and understandable to a tester \textit{across multiple task conditions} (inclusive of conditions that could invoke errors).
    \item[{[TrRL5]}] The system's performance is acceptable and understandable (traceable and logical) to an \textit{operator in a simulated environment}.
    \item[{[TrRL6]}] The system's performance is acceptable and understandable (traceable and logical) to an operator in a \textit{relevant environment}.
    \item[{[TrRL7]}] The system's performance is acceptable and understandable (traceable and logical) to an operator in an \textit{operational environment}.
    \item[{[TrRL8]}] The system's performance is acceptable and understandable (traceable and logical) to an operator \textit{across multiple task conditions} (inclusive of conditions known to invoke errors).
    \item[{[TrRL9]}] The system's performance is \textit{universally accepted and understood by the community of operators} across multiple task conditions (inclusive of conditions known to invoke errors).
\end{enumerate}
\section{Design Reference Mission}
\begin{figure}
    \centering
    \includegraphics[width = .4\textwidth]{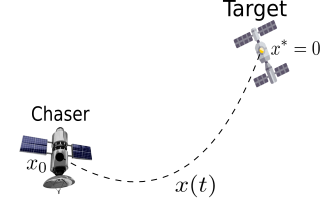}
    \caption{Autonomous Satellite Rendezvous and Docking}
    \label{fig:sat_docking}
\end{figure}
In this section, we consider the use case of an autonomous docking problem to exemplify the trusted autonomy levels. 
The first space docking maneuver was successfully performed in 1966 by Gemini 8 piloted by Neil Armstrong. In that mission, Neil manually piloted the Gemini spacecraft to rendezvous and dock with an uncrewed Agena Target Vehicle. 
In 1967, the first automated uncrewed docking, between Cosmos 186 and Cosmos 188, occurred by the USSR.  
Since the late 1960's, there has been a significant advancement to refine these maneuvers which makes it a prime test case for discussing the trusted autonomy readiness levels. These maneuvers have been done by hand, tele-operated, automated and autonomously. In 2020, researchers proposed a satellite docking challenge problem in \cite{PetersenSFM21}, with an under actuated satellite system. Numerous methods have been used to solve this problem in an algorithmic way considering different constraints and control methods \cite{SoderlundSciTech21a, SoderlundSciTech21b,Soderlund22SciTech}.

This example explores the hypothetical design of a neural network control system, trained using reinforcement learning for autonomous rendezvous, proximity operations, and docking (ARPOD). The docking scenario is depicted in Figure \ref{fig:sat_docking}, where a chaser spacecraft is approaching and docking with a target spacecraft. The chaser starts at an initial distance $x_0$, progresses in time to closer distances $x(t)$ and eventually reaches the target location $x^* = 0$. Hypothetical STAR Levels for the ARPOD scenario are as follows.


\begin{enumerate}
    \item[{[Ex. STAR Level 1]}] A company has decided to design ARPOD for satellite servicing, has identified that reinforcement learning might be the basis of this, and anticipates the solution will provide courses of action to operators to select from. The company builds a concept and considers design alternatives and risks.
    \item[{[Ex. STAR Level 2]}] It has been identified that an anticipated user is an operator who works in a ground control location for a spacecraft. Solutions will be measured based on fuel use and time to perform the mission. A number of potential safety concerns are identified (there are several risks associated with autonomous proximity operations \cite{hobbs2021risk,LangSciTech21}).  A prototype reinforcement learning solution is used to train a point mass model of a spacecraft to dock using linearized Clohessy-Wiltshire dynamics in Hill's reference frame \cite{dunlap2022run}. Rudimentary plots of performance are analyzed by researchers. 
    \item[{[Ex. STAR Level 3]}] A prototype black/gray box\footnote{A gray box system has some properties internal to the system known, while there is no requirement for a black box system to be known, \cite{doi:10.2514/6.2021-1126}.} advanced learning-based solution is developed for linearized three-dimensional Clohessy-Wiltshire dynamics and relative attitude dynamics. A need to bound the black/gray box solution with run time assurance algorithms is identified and potential prototypes are developed and compared \cite{dunlap2021comparing}. Feedback is received from operators on what parameters they might want to tweak in the docking control and run time assurance safety system. 
    \item[{[Ex. STAR Level 4]}] The solution graduates to higher fidelity digital twin spacecraft systems with modeled contact dynamics. Platform operators are identified and brought in to observe demonstrations. They provide constructive feedback of the scenarios on their concerns what kind of telemetry would like to see to assure safety approaches and acceptability of the solutions. Initial investigation on human-autonomy interface iterations between the platform solution and the operational community begins. 
    \item[{[Ex. STAR Level 5]}] The solution is demonstrated on a terrestrial spacecraft surrogate environment using space-like real time emulation platforms with operators interacting with the autonomy system prototype. Flight-like Processor in the loop computations are performed of the full ARPOD mission. Computational performance, memory footprint, and input quality and data rate requirements are identified.  A human-autonomy interface enables operators to understand the autonomy's behavior, decision rationale, and safety thresholds.
    \item[{[Ex. STAR Level 6]}] The spacecraft bus has been identified and the autonomous software and hardware ARPOD integrated  solution is installed on the actual hardware to verify performance within operationally realistic (real-time) constraints. The full system undergoes test computations with both nominal and off-nominal inputs representative of expected conditions of use.  A human-autonomy interface enables operators to understand test results in real time and provides them with feedback on safety, transparency, and performance.
    \item[{[Ex. STAR Level 7]}] The autonomous solution is launched onboard a spacecraft and monitored via the human-autonomy interface through a series of live flight tests. The live satellite on orbit practices its ARPOD strategy with a computer simulated spacecraft on the ground to evaluate potential scenarios with no risk of collision. Risks are quantified, and a strategy is defined for how data is obtained, managed, used, and secured. 
    \item[{[Ex. STAR Level 8]}] The behavior of the ARPOD system is tested in a variety of different scenarios on orbit while adjustments are made to the human-autonomy interface to satisfy spacecraft operators. The number of scenarios evaluated on the physical system is increasing. Two live spacecraft perform a stepwise progression through the phases of approach, proximity operations, and docking. For each phase the chaser spacecraft first computes what it would do, seeking permission from spacecraft operators before then performing that activity. Following successful docking, spacecraft operators undock and maneuver the chaser away, to repeat the scenario in different conditions (e.g., from different angles of approach, with illumination from different directions).
    \item[{[Ex. STAR Level 9]}] The novel neural network ARPOD system has been proven through successful mission operations for multiple programs and is continuously monitored and evaluated for anomalies in behavior. The accompanying run time assurance system has been demonstrated to consistently assure safety properties. The human-autonomy interface enables operators to successfully control the spacecraft, even in off-nominal conditions. 
\end{enumerate}

Hypothetical TrRLs for the ARPOD scenario are as follows.
\begin{enumerate}
    \item[{[TrRL1]}] The estimation and control approach for the ARPOD system's design seems reasonable to the designer. 
    \item[{[TrRL2]}] The ARPOD system's simulated performance is as expected to the designer.
    \item[{[TrRL3]}] The ARPOD's performance is acceptable in simulated tests. 
    \item[{[TrRL4]}] The ARPOD system's task performance is acceptable a tester across multiple task conditions.
    \item[{[TrRL5]}] The ARPOD system's performance as demonstrated on a terrestrial spacecraft surrogate environment is acceptable to an operator.
    \item[{[TrRL6]}] The ARPOD system's performance on the flight hardware is acceptable to operators.
    \item[{[TrRL7]}] The ARPOD system's performance is acceptable for use in a set of pre-planned tasks on orbit.
    \item[{[TrRL8]}] The ARPOD system's performance is acceptable to an operator in rendezvous from a variety of initial conditions as well as in a wide variety of tasks, such as formation flying, inspection, and docking.
    \item[{[TrRL9]}] The ARPOD system's performance is universally accepted, and requested by operators for use across multiple ARPOD task conditions.
\end{enumerate}




\section{Discussion}
After introducing autonomy and trust concepts, this research developed and illustrated a definition of space trusted autonomy readiness levels along axes of autonomy readiness levels and trust readiness levels. This effort is by no means the end of the definitions, rather it is a starting point for the purpose of wider community discussion on these topics. Readiness levels which accurately reflect the specific challenges in space trusted autonomy can lead to better program assessments and decisions, and accelerate technology adoption across the space community. We welcome feedback on these definitions, and are particularly interested in reports of their application (whether successful or problematic) to space autonomy, as a means to refine them further. Our lead author is the primary contact for such feedback.

\section*{Acknowledgements}
The Authors would like to thank Dr. Steve Rogers, Mr. JC Lede, Dr. Scott Clouse, Dr. Jared Culbertson, and Dr. Andrew Lacher.
A portion of this research was carried out at the Jet Propulsion Laboratory, California Institute of Technology, under a contract with the National Aeronautics and Space Administration (80NM0018D0004). The views expressed are those of the authors and do not reflect the official guidance or position of the United States Government, the Department of Defense, the United States Air Force, The National Aeronautics and Space Administration, the Jet Propulsion Laboratory, the National Reconnaissance Office, or the Aerospace Corporation.

\newpage
\section*{Appendix 1: Acronyms}
\begin{table}[htb!]
    \centering
    \label{tab:acronyms1}
    \begin{tabular}{p{1.2cm}p{6cm}}
        AEGIS & Autonomous Exploration for Gathering Increased Science\\
        AI &Artificial Intelligence\\
        ANGELS & Automated Navigation and Guidance Experiment for Local Space \\
        API & Application Programming Interface \\
        ARL & Algorithm Readiness Level \\
        ARPOD & Autonomous Rendezvous, Proximity Operations, and Docking\\
        ASTERIA & Arcsecond Space Telescope Enabling Research in Astrophysics \\
        C2BCM & Command Control Battle Management Communications\\
        CI/CD & Continuous Integration and Continuous Development\\
        CMMI & Capability Maturity Model Integration\\
        COTS & Commercial off-the-shelf\\
        CRL& Commercialization Readiness Level\\
        DAF & Department of the Air Force\\
        DARPA & Defense Advanced Research Projects Agency \\
        DART &Demonstration of Autonomous Rendezvous Technologies \\
        DRL & Data Readiness Level\\
        EDL & Entry, Descent and Landing\\
        EELV &Evolved Expendable Launch Vehicle \\
        ESA & European Space Agency\\
        ESPA & EELV Secondary Payload Adaptor\\
        FRP & Full Rate Production\\
        GPS & Global Positioning System\\
        HRL & Human Readiness Level\\
        LVC & Live-Virtual-Constructive\\
        LRIP & Low Rate Initial Production  \\
        ML & Machine Learning\\
        MLRL & Machine Learning Readiness Level \\
        MRL & Manufacturing Readiness Level\\
        MER & Mars Exploration Rover\\
        NASA & National Aeronautics and Space Administration\\
        NRO & National Reconnaissance Office\\
        P-LEO & Proliferated Low Earth Orbit\\
        R\&D & Research \& Development\\
        RAX & Remote Agent Experiment\\
        RL & Readiness Level\\
        RSGS & Robotic Servicing of Geosynchronous Satellites\\
        SAE & Society of Automation Engineers \\
        SDA & Space Domain Awareness \\
        SSN & Space Surveillance Network\\
        STAR  & Space Trusted Autonomy Readiness\\
        STM & Space Traffic Management \\
        SWaP & Size, Weight and Power \\
        TCL & Transition Commitment Level\\
        TRL & Technology Readiness Level\\
        TRN & Terrain Relative Navigation \\
        TrRL & Trust Readiness Level \\
        UAS & Unmanned Aircraft System \\
        USAF & United States Air Force \\
        USSF & United States Space Force \\
        V\&V & Verification and Validation \\
    \end{tabular}
\end{table}

\section*{Appendix 2: Alternative Readiness Levels}

The purpose of this manuscript is not to redefine TRLs, but rather to tailor them to the unique qualities of autonomous systems. For convenience, the standard TRLs are listed in Table \ref{tab:OG TRLs}.
\begin{table*}[htb]
    \centering
    \caption{Technology Readiness Levels \cite{AcqNotes2021TRL}}
    \label{tab:OG TRLs}
    \begin{tabular}{|p{0.8cm}|p{5cm}|p{10cm}|}\hline
Level &	Definition &	TRL Description\\\hline
1 &	Basic principles observed and reported.	Lowest level of technology readiness.  &Scientific research begins to be translated into applied research and development. Examples might include paper studies of a technology’s basic properties.\\\hline
2 &	Technology concept and/or application formulated. &	Invention begins. Once basic principles are observed, practical applications can be invented. Applications are speculative and there may be no proof or detailed analysis to support the assumptions. Examples are limited to analytic studies.\\\hline
3 &	Analytical and experimental critical function and/or characteristic proof of concept. &	Active research and development is initiated. This includes analytical studies and laboratory studies to physically validate analytical predictions of separate elements of the technology. Examples include components that are not yet integrated or representative.\\\hline
4 &	Component and/or breadboard validation in laboratory environment. &	Basic technological components are integrated to establish that they will work together. This is relatively “low fidelity” compared to the eventual system. Examples include the integration of “ad hoc” hardware in the laboratory.\\\hline
5 &	Component and/or breadboard validation in relevant environment. &	The Fidelity of breadboard technology increases significantly. The basic technological components are integrated with reasonably realistic supporting elements so it can be tested in a simulated environment.\\\hline
6 &	System/subsystem model or prototype demonstration in a relevant environment. &	A representative model or prototype system, which is well beyond that of TRL 5, is tested in a relevant environment. Represents a major step up in a technology’s demonstrated readiness.\\\hline
7 &	System prototype demonstration in an operational environment. &	Prototype near, or at, planned operational system. Represents a major step up from TRL 6, requiring the demonstration of an actual system prototype in an operational environment such as an aircraft, vehicle, or space.\\\hline
8 &	Actual system completed and qualified through test and demonstration. &	Technology has been proven to work in its final form and under expected conditions. In almost all cases, this TRL represents the end of true system development. Examples include developmental test and evaluations of the system in its intended system to determine if it meets design specifications.\\\hline
9 &	Actual system has proven through successful mission operations. &	The actual application of the technology in its final form and under mission conditions, such as those encountered in operational test and evaluation. Examples include using the system under operational mission conditions.\\\hline
    \end{tabular}
\end{table*}

This section includes paraphrased examples of different readiness levels from the literature including Algorithm RLs (ARLs)\cite{Bateman2016}, Manufacturing RLs (MRLs)\cite{morgan2015}, Commercialization RLs (CRLs)\cite{Maybury2014}, Machine Learning RLs (MLRLs) \cite{lavin2021technology}, Human RLs (HRLs)\cite{ANSI_HFES400_2021}, and Transition Commitment Levels (TCLs) \cite{TCLs}. Data RLs (DRLs)\cite{lawrence2017data} are considered separately because a 1-9 scale was not used. Note that many of these definitions are much more extensive than the summaries presented below. In particular MLRLs have long descriptions of each level, which are summarized to a few sentences each here that may lose nuance. 

\subsection{Level 0}
Only MLRLs included a level zero.
\begin{enumerate}
    \item[{[MLRL0]}] \textbf{First Principles}: consists of literature review, building mathematical foundations, white-boarding concepts and algorithms, and building an understanding of the data requirements; no data available; reviewed by research lead, such as a PhD supervisor, for mathematical validity and potential novelty or utility 
\end{enumerate}

\subsection{Level 1}
\begin{enumerate}
    \item[{[TRL1]}] Basic principles observed and reported.
    \item[{[ARL1]}] Need identified, analysis in any suitable format, input data sufficient to demonstrate basic operation.
    \item[{[MRL1]}] Basic manufacturing implications identified. 
    \item[{[CRL1]}] {Research and Development}: problem definition, potential business/stakeholder interest, preliminary business case (market, margin/savings) and baseline, investment from external partner, intellectual property disclosure.
    \item[{[MLRL1]}] {Goal-Oriented Research}: analyze specific model or algorithm properties; collect and process sample (possibly synthetic) data for model evaluation and training; research-caliber code; start versioning of code, models, datasets; review by research team members with many iterations of feedback and experiments. 
    \item[{[HRL1]}] Basic principles for human characteristics, performance, and behavior observed and reported. 
    \item[{[TCL1]}] Internal program/R\&D commitment.
\end{enumerate}


\subsection{Level 2}
\begin{enumerate}
    \item[{[TRL2]}] Technology concept and/or application formulated. 
    \item[{[ARL2]}] Concepts identified, analysis in any suitable format, input data sufficient to demonstrate basic operation. 
    \item[{[MRL2]}] Manufacturing concepts identified. 
    \item[{[CRL2]}] Core business interest:  business/stakeholder engagement strategy, initial transition plan, accountable core business champion, refined business case and baseline - realistic early stage intellectual property evaluation.
    \item[{[MLRL2]}] Active R\&D is initiated with simulation environments or simulated data. An initial requirements document states model-specific technical goals, and verification and validation steps. Data sets may be publicly available, semi-simulated, or fully simulated. 
    \item[{[HRL2]}] Human-centered concepts, applications, and guidelines defined. 
    \item[{[TCL2]}] Internal portfolio commitment.
\end{enumerate}

\subsection{Level 3}
\begin{enumerate}
    \item[{[TRL3]}] Analytical and experimental critical function and/or characteristic proof of concept.
    \item[{[ARL3]}] Proof of Concept, analysis in any suitable format, input data sufficient to demonstrate basic operation.
    \item[{[MRL3]}] Manufacturing proof of concept 
    \item[{[CRL3]}] Interaction/Awareness - periodic, planned business/stakeholder engagement.
    \item[{[MLRL3]}] Transition from research code to robust and clean prototype-caliber code that emphasizes interoperability, reliability, maintainability, extensibility, and scalability that is well-designed, well-architected for dataflow and interfaces, generally covered by unit and integration tests, meet team style standards, and sufficiently-documented. A cross-disciplinary team from applied AI and engineering review software practices, interfaces and documentation, version control for models and datasets, and domain- or organization-specific data management considerations. 
    \item[{[HRL3]}] Human-centered requirements to support human performance and human technology interactions established. 
    \item[{[TCL3]}] Sponsor/customer interaction and awareness.
\end{enumerate}

\subsection{Level 4}
\begin{enumerate}
    \item[{[TRL4]}] Component and/or breadboard validation in a laboratory environment.
    \item[{[ARL4]}] Prototype developed, standalone component, \mbox{Matlab} analysis, input data sufficient to demonstrate basic operation.
    \item[{[MRL4]}] Manufacturing processes in lab environment.
    \item[{[CRL4]}] Commitment/support - access to business/stakeholder information, users, customers, environment; realistic early stage intellectual property valuation; market/pricing distribution plan; business/stakeholder investment. 
    \item[{[MLRL4]}] A quick proof-of concept example is developed using representative data to explore candidate applications and provide results for qualitative and quantitative analysis such as model and algorithm performance (e.g., precision and recall and various data splits), computational costs (e.g., CPU vs GPU runtimes), and also metrics that are more relevant to the eventual end-user (e.g., number of false positives).
    \item[{[HRL4]}] Modeling, part-task testing, and trade studies of human systems design concepts and applications completed. 
    \item[{[TCL4]}] Sponsor/customer commitment and active support.
\end{enumerate}

\subsection{Level 5}
\begin{enumerate}
    \item[{[TRL5]}] Component and/or breadboard validation in a relevant environment.
    \item[{[ARL5]}] Reference implementation in closed loop environment, component/capability integrate in reference implementation but other components may not be updated, Command Control Battle Management Communications (C2BCM) benchmark, and some stressing vignettes recorded flight test/ground test data for components not requiring sensor/element interaction.
    \item[{[MRL5]}] Components in production relevant environment. 
    \item[{[CRL5]}] Pilot Plan - business/stakeholder commitment to direct funded operational pilot using operational data and environment; environment set up; business model articulated; success measures and criteria defined. 
    \item[{[MLRL5]}] A specific machine learning capability is demonstrated and evaluated by an audience beyond ML researchers, possibly with example scripts and/or an API. This stage also represents transition from an R\&D prototype team to a product development team, and the beginning of the ``value of death.” Considerations are made for data scaling and governance.
    \item[{[HRL5]}] Human-centered evaluation of prototypes in mission-relevant part-task simulations completed to inform design.
    \item[{[TCL5]}] Sponsor/customer commitment to pilot.
\end{enumerate}

\subsection{Level 6}
\begin{enumerate}
    \item[{[TRL6]}] System/subsystem model or prototype demo in a relevant environment. 
    \item[{[ARL6]}]  Reference implementation system validation, Master reference implementation - all components updated for the spiral, Command Control Battle Management Communications (C2BCM)benchmark, and stressing vignettes.
    \item[{[MRL6]}] System or subsystem in production relevant environment. 
    \item[{[CRL6]}] Operational Pilot: business-funded operational pilot; results analysis.
    \item[{[MLRL6]}] Transition from prototype code to product-caliber machine learning code with precise specifications, test coverage, and well-defined APIs. Considerations are made to design model explanations for stakeholders rather than ML engineers. Focus is on the code quality, an AI ethics revisit, regulatory compliance, data privacy and security laws. 
    \item[{[HRL6]}] Human systems design fully matured as influenced by human performance analyses, metrics, prototyping, and high fidelity simulations. 
    \item[{[TCL6]}] Sponsor/customer execution of operational pilot.
\end{enumerate}

\subsection{Level 7}
\begin{enumerate}
    \item[{[TRL7]}] System prototype demo in an operational environment. 
    \item[{[ARL7]}] Developed in system, all components for spiral updated and integrated into intended operational software, Command Control Battle Management Communications (C2BCM) analysis, Verification Scenarios and parametric analysis.
    \item[{[MRL7]}] System or subsystem demonstrated in production representative environment. 
    \item[{[CRL7]}] Purchased: business/stakeholder capability acquisition plan; benefit measures established.
    \item[{[MLRL7]}] The ML technology is integrated into a system by a multidisciplinary team with expertise beyond AI/ML and tested for specific use case scenarios. Risks are quantified, and a strategy for how data is obtained, managed, used, secured, and ethically used is developed by quality assurance engineers.
    \item[{[HRL7]}] Human systems design fully tested and verified for a range of scenarios and tasks in operational environment with system hardware, software, and representative users.
    \item[{[TCL7]}] Sponsor/customer commitment
to acquisition.
\end{enumerate}

\subsection{Level 8}
\begin{enumerate}
    \item[{[TRL8]}] Actual system completed and qualified through test and demo. 
    \item[{[ARL8]}] Fielded in System, Command Control Battle Management Communications Algorithm and Analysis Environment, and flight test/ground test scenarios. 
    \item[{[MRL8]}] Pilot Line demonstrated ready for low rate initial production (LRIP). 
    \item[{[CRL8]}] Profit/Savings: Measured capability benefit achieved by a single business/stakeholder.
    \item[{[MLRL8]}] Flight-ready technology is demonstrated to work in its final form in anticipated conditions. The system has been stress tested during continuous integration and continuous development (CI/CD) processes, and additional tests such as A/B tests, blue/green deployment tests, shadow testing, and canary testing are conducted. When using real data, the ML may be tested in a shadow mode first to evaluate performance degradation. A panel representing a full set of stakeholders makes a go or no-go decision for deployment. 
    \item[{[HRL8]}] Total human-system performance fully tested, validated, and approved in mission operations, using completed system hardware, software. and representative users. 
    \item[{[TCL8]}] Mission impact realized.
\end{enumerate}

\subsection{Level 9}
\begin{enumerate}
    \item[{[TRL9]}] Actual system proven through successful mission operations.
    \item[{[ARL9]}] Mature, Command Control Battle Management Communications Algorithm and Analysis Environment, and Live Scenarios.
    \item[{[MRL9]}] LRIP demonstrated, ready for full rate production (FRP). 
    \item[{[CRL9]}] At Scale: additional business/stakeholders or domains identified; measured benefit (profit/savings/capability) achieved across domains.
    \item[{[MLRL9]}] AI and ML technologies are deployed, there is continuous monitoring and evaluation for performance and reliability degradation, and identification of explicit considerations for improving the next version. Data logs are used to capture inputs, model predictions, and anomalies or deviations in model behavior. There is a path for user communication and feedback to the R\&D team. Any changes to the system or components based on continuous monitoring or user feedback revert the system to TRL 7. 
    \item[{[HRL9]}] System successfully used by the intended users in operations across the operational envelope with systematic monitoring of human system performance issues, errors, and accidents to identify enhancements. 
    \item[{[TCL9]}] Impact scaled out.
\end{enumerate}

\subsection{Level 10}
\begin{enumerate}
    \item[{[MRL10]}] Full rate production (FRP) demonstrated, lean production practices in place. 
\end{enumerate}

\section*{Appendix 3: Space Autonomy Missions}
There is a long history in autonomous and artificial intelligent operations throughout the space domain. Incremental approaches to mature technology along TRLs have been documented, including the buildup of missions in preparation for the moon, or rapid advancement of space technologies under NASA's New Millennium Program \cite{minning2003technology}.
This section provides context with examples of major satellite and spacecraft experimental programs to test autonomy technologies in space.

\subsection{Deep Space One}
The first use of AI in space took place in 1998, when NASA launched the Deep Space One spacecraft on a mission to validate in space several high-risk, new technologies important for future space programs \cite{rayman2000results}.  Three autonomy technologies were tested: beacon monitoring determined the overall spacecraft health from on-board data and transmitted that status back to Earth, \textit{AutoNav} used its determined position and velocity to compute and execute maneuvers to deliver the spacecraft to its target, and the Remote Agent Experiment (RAX) that represented the first AI agent to control a NASA spacecraft without human supervision. The RAX demonstrated autonomous operations of the craft in two scenarios of 18 hours and 6 hours respectively, achieving all its validation objectives. 
In a hierarchical fashion, RAX was given high level objectives (or commander's intent) and was allowed to derive its own primitive maneuvers to achieve the objective. It was successful in demonstrating the ability to plan, diagnose and respond to simulated faults in the spacecraft. 

\subsection{Mars Rovers: Testing Autonomy During Extended Mission Phase}
As an alternative to launching an entire mission devoted to flight validation, autonomy testing is sometimes done during an extended mission phase, after the primary mission has been accomplished and the ground team is more willing to accept the risk of employing a novel technology.
This was the case for the Autonomous Exploration for Gathering Increased Science (AEGIS) system, which enables automated data collection by planetary rovers. It was first uploaded to the Mars Exploration Rover (MER) mission’s Opportunity rover in December 2009, by which time the rover had been operating on Mars for almost six years. AEGIS successfully demonstrated autonomously selecting targets based on scientist-specified objectives in images from the rover’s navigation camera and acquiring follow-up images with the narrow field of view science camera 
\cite{estlin2012aegis}. In a similar manner, AEGIS was later deployed to NASA’s Curiosity Mars rover, where it has been in routine use since May 2016, selecting targets for the ChemCam remote geochemical spectrometer instrument \cite{ francis2017aegis}. These two successful applications led to AEGIS being included in the primary mission operations of NASA’s latest Mars rover, Perseverance.

\subsection{ASTERIA CubeSat: Testing Autonomy During Extended Mission Phase}
Another example of use of mission extension for testing was the Earth-orbiting Arcsecond Space Telescope Enabling Research in Astrophysics (ASTERIA) CubeSat, which used its extended mission to demonstrate autonomy technologies, including a shift of the spacecraft commanding paradigm from time-based sequence to Task Networks, and separately, demonstration of on-board orbit determination in Low Earth Orbit without the Global Positioning System (GPS). Further demonstrations were planned but instead were run on the testbed following the CubeSat’s demise \cite{fesq2021results}.


\subsection{Autonomous Optical Navigation for Interplanetary Missions}
Turning spacecraft navigation into an on-board completely autonomous capability has the benefit of reducing the Earth-based operational cost and effort, and enables missions that depend on critical maneuvers in situations where the round-trip light-time precludes ground-in-the-loop interaction. Historically, navigation of deep space missions has been controlled from Earth. Doppler and range tracking via radio contact with the spacecraft, combined with optical imagery taken by the spacecraft and relayed to Earth, have provided the information to compute the spacecraft’s trajectory and trajectory correction maneuvers as needed. 
Optical data is well suited to being acquired and processed autonomously to form the basis for a completely autonomous navigation system. This was demonstrated on NASA’s Deep Space One mission \cite{ bhaskaran1996autonomous}, and subsequently used successfully on camera-equipped NASA’s comet fly-bys (comets Borrelly, Wild 2, Tempel 1, and Hartley 2), an asteroid flyby (Annefrank) \cite{ bhaskaran2012autonomous }, and for the trajectory control of the Deep Impact mission’s impactor to ensure it would hit the nucleus of its target, comet Tempel 1 \cite{ kubitschek2006deep}.

\subsection{Autonomous Optical Navigation for Orion Artemis Missions}
As part of NASA’s Orion Artemis Missions, OpNav (optical navigation) has been developed to be the first autonomous safety-critical on-board navigation system, able to complete the mission if Earth communication is lost.  Its on-board computed navigation data would substitute for nominally provided ground trajectory course maneuver updates. Its development followed a traditional waterfall software development lifecycle with periodic major reviews, progressing from TRL 3 to TRL 8 while adhering to the NASA Software Engineering Requirements standard, NASA NPR 7150.2B, and the practices of the Capability Maturity Model Integration (CMMI) Level 3 \cite{prokop2021case}. Formal validation and certification of OpNav is planned during the outbound leg of the Artemis I mission, to bring it to TRL 9.

\subsection{Autonomous Entry, Descent and Landing} The landing of robotic spacecraft on Mars exemplifies a situation where on-board control is essential. Colloquially referred to as the “seven minutes of terror,” the entire EDL sequence is over in less than the one-way light time to communicate with the craft. From the first successful soft landing on Mars, achieved by the Soviet Mars 2 probe, this has been done automatically. Autonomy was introduced on the landings of the two Mars Exploration Rovers (MERs) via the Descent Image Motion Estimation System (DIMES), the first ever terrain-relative sensing and guidance system used by a real space mission \cite{ johnson2007design}. More recently, Terrain Relative Navigation (TRN), was used in 2021 to guide the pin-point landing of NASA’s 2020 Mars rover mission. This enabled the mission to aim for Jezero Crater yet avoid many hazards therein (cliffs, dune fields, rocks), resulting in a landing within 5m of the targeted location \cite{ johnson2022mars}.  In \cite{ beauchamp2020technology}, the development of TRN serves as an illustration of progression through the nine NASA Technology Readiness Levels, from which the following summary is distilled.
\begin{enumerate}
\item TRL 1: Pinpoint landing concepts explored
\item TRL 2: Benefits and desired performance characteristics identified
\item TRL 3: Performance analyses and experimental proof-of-concept using descent imagery from previous Mars landings
\item TRL 4: Algorithms tested off-line, including data from a sounding rocket flight emulating a Mars landing
\item TRL 5: Performance shown on prototype computing hardware connected with  Commercial off-the-shelf (COTS) hardware
\item TRL 6: Real-time tests of implementation over a wide variety of scenes, gathered on helicopter flights
\item TRL 7: Demonstrated on a vertical take-off and vertical landing rocket
\item TRL 8: Implementation completed, environmentally tested, and delivered for spacecraft integration
\item TRL 9: Successful use on Mars 2020 rover mission’s landing 
\end{enumerate}

\bibliographystyle{IEEEtran}
\bibliography{references}

\thebiography
\begin{biographywithpic}{Kerianne L. Hobbs}{Figures/HobbsKerianne8x10}
is the Safe Autonomy and Space Lead on the Autonomy Capability Team (ACT3) at the Air Force Research Laboratory. There she investigates rigorous specification, analysis, bounding, and intervention techniques to enable safe, trusted, ethical, and certifiable autonomous and learning controllers for aircraft and spacecraft applications. Her previous experience includes work in automatic collision avoidance and autonomy verification and validation research. Dr. Hobbs was selected for the 2020 AFCEA 40 Under 40 award and was a member of the team that won the 2018 Collier Trophy (Automatic Ground Collision Avoidance System Team), as well as numerous AFRL Awards.  Dr. Hobbs has a BS in Aerospace Engineering from Embry-Riddle Aeronautical University, an MS in Astronautical Engineering from the Air Force Institute of Technology, and a Ph.D. in Aerospace Engineering from the Georgia Institute of Technology. 
\end{biographywithpic}

\begin{biographywithpic}{Joseph B. Lyons}{Figures/JoeLyons}
is a Principal Research Psychologist within the 711 Human Performance Wing at Wright-Patterson AFB, OH. Dr. Lyons received his PhD in Industrial/Organizational Psychology from Wright State University in Dayton, OH, in 2005. Some of Dr. Lyons’ research interests include human-machine trust, interpersonal trust, human factors, and influence. Dr. Lyons has worked for the Air Force Research Laboratory as a civilian researcher since 2005, and between 2011-2013 he served as the Program Officer at the Air Force Office of Scientific Research where he created a basic research portfolio to study both interpersonal and human-machine trust as well as social influence. Dr. Lyons has published in a variety of peer-reviewed journals, and is an Associate Editor for the journal Military Psychology. Dr. Lyons is an AFRL Fellow, Fellow of the American Psychological Association, and Fellow of the Society for Military Psychologists. Dr. Lyons can be contacted at: joseph.lyons.6@us.af.mil.
\end{biographywithpic}

\begin{biographywithpic}{Martin S. Feather}{Figures/MartinFeather}
is a Principal Software Assurance Engineer in JPL’s Office of Safety and Mission Success. His focus is on research to assure space missions, in particular their software. A recipient of a NASA Exceptional Achievement Medal, he has been an author on over 180 publications spanning a range of topics, with the common theme of viewing the assurance problem from the perspective of what to be concerned about, and how to show those concerns are either absent or adequately addressed. He received his BA and MA in Mathematics and Computer Science from the Cambridge University, UK, and PhD in Artificial Intelligence from the University of Edinburgh, UK.
\end{biographywithpic}

\begin{biographywithpic}{Benjamen P. Bycroft}{Figures/BenBycroft} is a Senior 
Vehicle Engineering and Autonomy Project Leader at The Aerospace Corporation. His work spans research, development, and application of advanced autonomy, dynamics and control, as well as electromechanical systems engineering for both launch vehicles and spacecraft. Benjamen has previously taught courses within the University of Southern California’s Viterbi School of Engineering. He holds a B.S. in Mechanical Engineering and an M.S. in Dynamics \& Control from USC, as well as a B.A. in Physics from Willamette University.\end{biographywithpic}

\begin{biographywithpic}{Sean Phillips}{Figures/Phillips_img} is the Deputy Technical Advisor for the Space Control Branch, Project Lead for the Local Intelligent Networked Collaborative Satellites (LINCS) Lab and a Research Mechanical Engineer at the Air Force Research Laboratory, Space Vehicles Directorate. He is a Research Assistant Professor (LAT) at the University of New Mexico in Albuquerque, NM. He received his Ph.D in the Department of Computer Engineering at the University of California – Santa Cruz in 2018. He received his B.S. and M.S. in Mechanical Engineering from the University of Arizona in 2011 and 2013, respectively. Dr. Phillips current research interests consist of robust GNCA algorithm development, hybrid system modeling and analysis, satellites systems, communication/information network, and distributed systems under adverse conditions. \end{biographywithpic}

\begin{biographywithpic}{Michelle Simon}{Figures/MichelleSimon} is the technical lead for the Autonomous Operations Resilience for Tactile Action Program. She been with Air Force Research Laboratory since 2016 and has over 10 years in the aerospace industry. She currently works in the Space Vehicles Department located in Albuquerque, NM where her main area of research is autonomous systems. She received a B.S. in Computer Programming from Pacific Lutheran University in Tacoma, WA in 2010 and an M.S. in Electrical Engineering from University of Alaska in Fairbanks, AK in 2016.
\end{biographywithpic}

\begin{biographywithpic}{Mark Harter}{Figures/MarkHarter} is a Principal Space Systems Engineer leading several major space projects at The MITRE Corporation in Colorado Springs, CO. Over the last 16 years he has supported space
projects with the U.S. Space Force (USSF), United States Space Command (USSPACECOM), Missile Defense Agency (MDA), and Office of Space Commerce (OSC). Prior to MITRE, Mr. Harter was a Space and Missile Operations Officer in the Air Force for 23 years with experience in Space Control, Global Positioning System (GPS), Space Weapons \& Tactics, Nuclear Command \& Control, Minuteman III ICBMs, and Hypersonic systems. He has a B.S. in Aerospace Engineering, MS in Business, and MS in Military Operations.
\end{biographywithpic}  

\begin{biographywithpic}{Kenneth Costello}{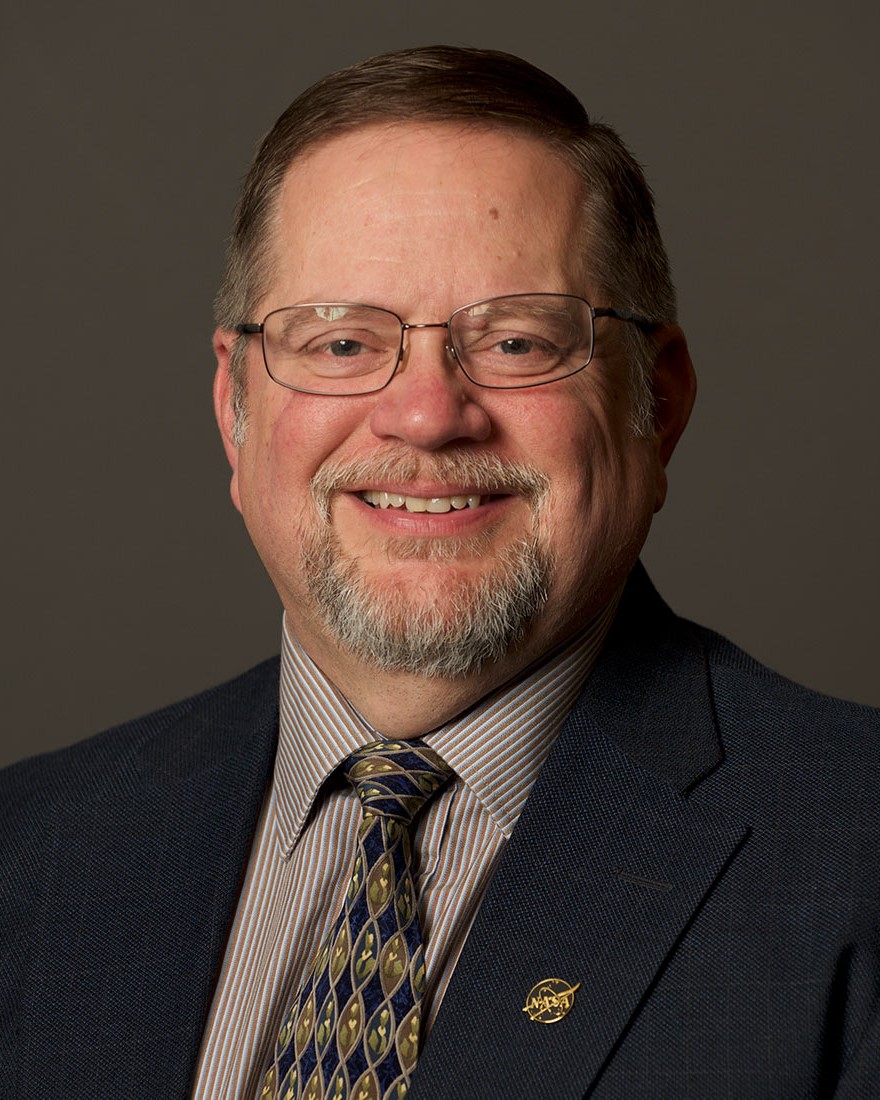} is the lead for the Engineering Services Group in the Safety and Mission Support Office of NASA’s Independent Verification and Validation (IVV) Program. The group provides various engineering services from software safety assessments to cybersecurity assessments as well as supporting the IVV Program in developing new analysis techniques. Ken has over 30 years of experience in software engineering and testing and holds a B.S. in Aerospace Engineering from the Pennsylvania State University and a M.S. in Software Engineering from West Virginia University where he also teaches software engineering classes.
\end{biographywithpic} 

\begin{biographywithpic}{Yuri O. Gawdiak}{Figures/YGImage} has a B.S. from CMU in Information Systems \& Ergonomics. He has worked at the Library of Congress, Fairchild, the Defense Communication Agency, the Univ. of Maryland HCI Lab, and at Boeing on the Space Station program. At NASA Yuri was a PI on STS-76/Mir-21 and supported payload avionics for Space Station. At ARC Mr. Gawdiak was a project manager in the Aviation Safety and in the Airspace Systems Programs, build manager for the Surface Movement Advisor at Atlanta Hartsfield International Airport, lead the PSA Space Station robot research, and director for Engineering for Complex Systems program. Mr. Gawdiak was also the portfolio management director at the Joint Planning \& Development Office on the multiagency NextGen effort and developed and proposed On Demand Mobility (ODM) as way to address aviation’s emissions challenges. Currently, Yuri is the associate director for Airspace Operations \& Safety Program, supports the Advanced Air Mobility Integration Office, as well as a member on various agency level teams on model based systems engineering, risk management, and ethical AI/ML development. Mr. Gawdiak has received both the NASA Exceptional Achievement \& Outstanding Leadership Medals.
\end{biographywithpic}

\begin{biographywithpic}{Lt Col Stephen Paine}{Figures/StevePaine} leads research in Synthetic Cognition for the National Reconnaissance Office’s Advanced Systems and Technology Directorate, exploring autonomous behavior in space operations.  He is a U.S. Air Force developmental engineer with extensive experience applying novel techniques in satellite ground system development and operations.  Lt Col Paine holds a BS in Electrical Engineering from the US Air Force Academy and MSEE in Space Systems Engineering from University of Colorado, Colorado Springs.\end{biographywithpic}


\end{document}